\documentclass[letterpaper,aps,prl,twocolumn,showcaps,showpacs,superscriptaddress,amsmath,amssymb]{revtex4}
\usepackage{graphicx,color,longtable}
\usepackage{amsmath,amsfonts,amsthm} 		
\usepackage{epstopdf}
\usepackage[caption=false]{subfig}

\definecolor{blue(ncs)}{rgb}{0.0, 0.53, 0.74}

\newcommand{\paddyspeaks}[1]{{\color{black} #1}}

\begin{document}

\makeatletter

\title{Experimental Evidence for a Structural-Dynamical Transition in Trajectory Space}

\author{Rattachai Pinchaipat}
\affiliation{H.H. Wills Physics Laboratory, Tyndall Avenue, Bristol, BS8 1TL, UK}
\affiliation{Centre for Nanoscience and Quantum Information, Tyndall Avenue, Bristol, BS8 1FD, UK}

\author{Matteo Campo}
\affiliation{Graduate School Materials Science in Mainz, Staudinger Weg 9, 55128 Mainz, Germany}
\affiliation{Institut f\"ur Physik, Johannes Gutenberg-Universit\"at Mainz,  Staudingerweg 7-9, 55128 Mainz, Germany}

\author{Francesco Turci}
\affiliation{H.H. Wills Physics Laboratory, Tyndall Avenue, Bristol, BS8 1TL, UK}
\affiliation{Centre for Nanoscience and Quantum Information, Tyndall Avenue, Bristol, BS8 1FD, UK}

\author{James E. Hallett}
\affiliation{H.H. Wills Physics Laboratory, Tyndall Avenue, Bristol, BS8 1TL, UK}
\affiliation{Centre for Nanoscience and Quantum Information, Tyndall Avenue, Bristol, BS8 1FD, UK}

\author{Thomas Speck}
\affiliation{Institut f\"ur Physik, Johannes Gutenberg-Universit\"at Mainz,  Staudingerweg 7-9, 55128 Mainz, Germany}

\author{C. Patrick Royall}
\affiliation{H.H. Wills Physics Laboratory, Tyndall Avenue, Bristol, BS8 1TL, UK}
\affiliation{Centre for Nanoscience and Quantum Information, Tyndall Avenue, Bristol, BS8 1FD, UK}
\affiliation{School of Chemistry, University of Bristol, Cantock's Close, Bristol, BS8 1TS, UK}
\affiliation{Department of Chemical Engineering, Kyoto University, Kyoto 615-8510, Japan}

\begin{abstract}
Among the key insights into the glass transition has been the identification of a non-equilibrium phase transition in \emph{trajectory space} which reveals phase coexistence between the normal supercooled liquid (active phase) and a glassy state (inactive phase). Here we present evidence that such a transition occurs in experiment. In colloidal hard spheres we find a non-Gaussian distribution of trajectories leaning towards those rich in locally favoured structures (LFS), associated with the emergence of slow dynamics. This we interpret as evidence for an non-equilibrium transition to an inactive LFS-rich phase. Reweighting trajectories reveals a first-order phase transition in trajectory space between a normal liquid and a LFS-rich phase. We further find evidence of a purely dynamical transition in trajectory space.
\end{abstract}

\pacs{64.70.Q-; 64.70.P-; 61.20.Gy; 61.20.-p}


\maketitle

\textit{Introduction. --- }
The glass transition is one of the longstanding challenges in condensed matter. In particular, one seeks to understand how solidity emerges with little apparent change in structure \cite{berthier2011}. A central aspect for the understanding of supercooled liquids is dynamic heterogeneity: on suitable observation time scales, local regions appear liquid-like (\emph{active}) or solid-like (\emph{inactive}) \cite{perera1996}, suggesting that any successful explanation must include this phenomenon. A variety of theories have been proposed, indeed whether the glass transition has a thermodynamic (implying structural) or dynamical origin remains unclear~\cite{berthier2011}. The former may relate to a transition to an \emph{ideal glass} state at finite temperature with minimal configurational entropy and has recently received some support from numerical and theoretical work ~\cite{cammarota2012,berthier2013overlap,ozawa2015}.

The dynamical interpretation posits that the glass transition is a dynamical phenomenon where local relaxation events in the form of active regions couple to one another~\cite{chandler2010}. This \textit{dynamic facilitation} approach employs the language of phase transitions in order to explain the emergence of solidity, but with a key departure from equilibrium thermodynamics: here the phase transitions occur in \emph{trajectory space}~\cite{jack2007,hedges2009,thompsonthesis} instead of configurational space. In such transitions, trajectories of small systems, of the duration of a few structural relaxation times, exhibit a transition between active and inactive states under a biasing field, which in simulation is compatible with the scaling expected for a first-order transitions~\cite{hedges2009,speck2012jcp}. It is suggested that the dynamical heterogeneity exhibited by glassforming liquids is the hallmark of such a dynamical phase transition~\cite{chandler2010}.

An extension of this trajectory space approach concerns structural-dynamical transitions which may provide a link between the thermodynamic (structure-based) and dynamical transition approaches~\cite{speck2012,turci2016}. Here one exploits the fact that, while changes in structure upon supercooling in liquids are not dramatic, nor are they absent \cite{royall2015physrep}. In fact, they contribute to the emergence of strongly heterogenous dynamical states. In particular, for a variety of model glassformers, geometric motifs known as locally favored structures (LFS) associated with slow dynamics have been identified~\cite{royall2014}.

This suggests that the dynamical phase transition in trajectory space may have a structural element, with the inactive phase having exceptionally high concentrations of LFS with respect to the active phase. This has been shown to be the case~\cite{speck2012}. Moreover selecting trajectories rich in LFS (rather than being dynamically inactive) leads to a similar non-equilibrium phase transition between a glassy LFS-rich phase and the normal (LFS-poor) supercooled liquid.  
This transition was found by biasing the population of LFS \emph{along the length of a trajectory}. The effect of the bias amounts to a dynamical chemical potential for the time-averaged LFS population favouring the sampling of trajectories rich or poor in structure. To realise the non-equilibrium transition, in practise a field termed $\mu$ is applied which uses a Boltzmann weight to sample trajectories based on their time-averaged LFS population. This demonstrates coupling between structure and dynamics.

Now to date, dynamical transitions have been carried out under such biasing fields, which are of course absent in experiment. However, even in equilibrium, evidence of such transitions can be found by considering the so-called large deviations of dynamical observables which serve as order parameters for the transition~\cite{touchette2010}. Quantities such as the mobility~\cite{hedges2009} or, the time-integrated population of LFS~\cite{speck2012}, exhibit non-Gaussian probability distributions with enhanced tails corresponding to exceptionally large (or small) values of the observable. Non-convexity of these distributions indicates a non-equilibrium phase transition which is revealed by reweighting these distributions (equivalent to applying the dynamical chemical potential) in the form of two coexisting peaks in the distribution of the observable of interest averaged along the trajectory. In experiments at equilibrium, a correct sampling identification of the non-Gaussian tails thus indicates the transition.

Particle-resolved studies of colloids~\cite{ivlev} provide data similar to that of computer simulation, and have been used to show structural change approaching dynamical arrest \cite{leocmach2012,mazoyer2011,royall2008} and that shear banding may be interpreted as a non-equilibrium transition~\cite{chikkadi2014}. Moreover simulation data show behaviour consistent with dynamical facilitation \cite{isobe2016}. Our aim here is to seek an \emph{experimental} signature of the dynamical phase transitions in time-averaged LFS populations and mobility, which we confirm with computer simulations. To do this we apply the $\mu$ field as post-processing to the experimentally determined non-Gaussian distributions. We back up our results with simulations in two ways. Firstly, we employ biased sampling using small systems similar to that used in~\cite{speck2012}. We then use larger, unbiased, simulations which we \emph{subsample} to obtain trajectories corresponding to a small system, and show that the transition is accessible to experiment.

\begin{figure}
\centering
\includegraphics[width=85mm]{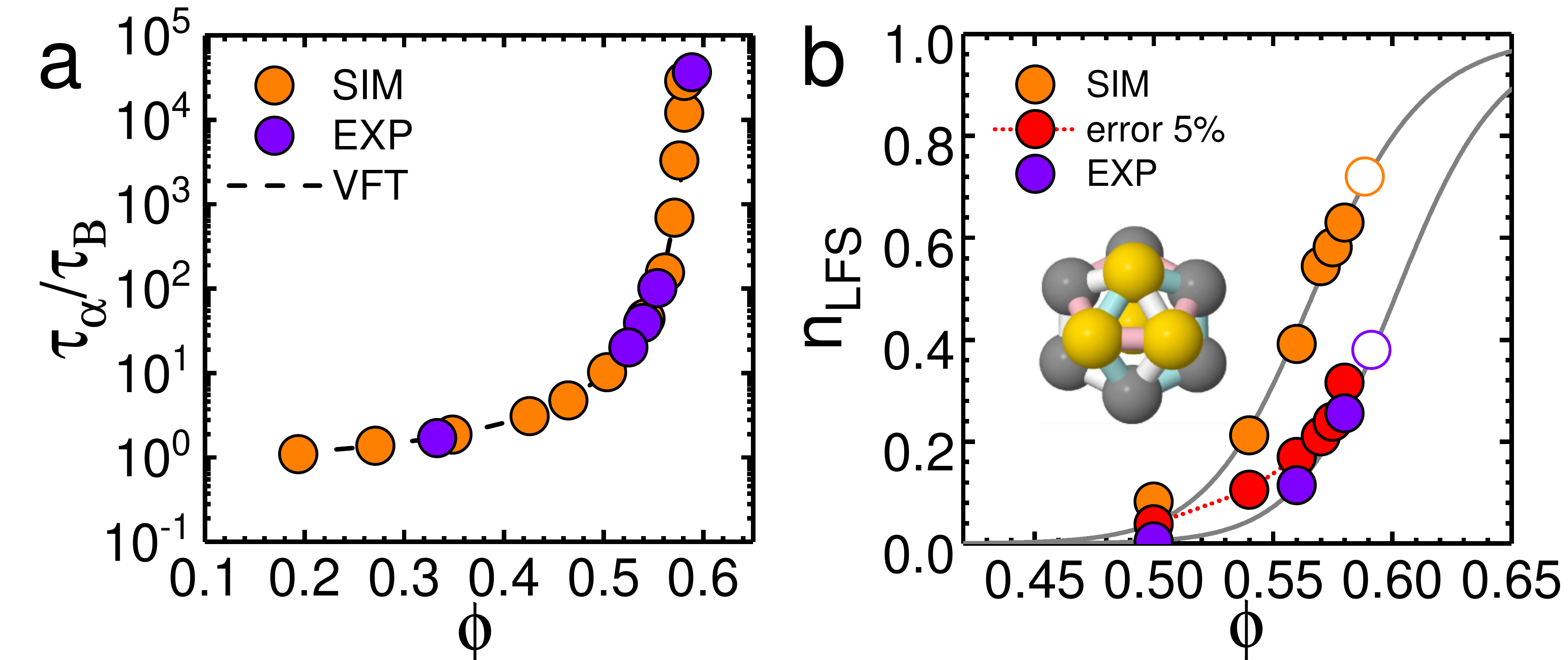}
\caption{Dynamical behaviour and structural changes upon supercooling hard spheres. (a) Angell plot of structural relaxation time $\tau_\alpha$ as a function of 
volume fraction. The dashed line is the VFT fit as described in the text. 
(b) Fraction of particles identified in defective icosahedra locally favoured structures $n_\mathrm{LFS}$ increases upon supercooling. Simulation data with errors added to the coordinates (red symbols) shows quantitative agreement with experiment. Unfilled symbols indicate volume fraction corresponding to LFS population in LFS-rich phase. Grey lines are fits to $n_\mathrm{LFS}(\phi)$ (see SM \cite{SM}.)}
\label{figAngell}
\end{figure}

\textit{Experiment. --- } 	
We used poly(methyl methacrylate) (PMMA) colloids fluorescently labelled with mean diameter of $\sigma=1.99$ $\mu$m and polydispersity 8$\%$. 
The particles were suspended in a density matched solvent to which salt was added to screen electrostatic interactions. We use confocal microscopy to track the particle coordinates~\cite{Leocmach2015}. Due to particle tracking limitations errors are introduced in the coordinate data \cite{royall2013myth,royall2007jcp}. To determine the impact of the errors we compared experiments with simulation as shown in Fig. \ref{figAngell}(b). Here we see that, applying a Gaussian distributed error with standard deviation $0.05\sigma$ to the simulation data leads to comparable results to the experiments. Further details may be found in the Supplementary Material (SM) \cite{SM}.

\begin{figure*}
\includegraphics[width=\linewidth]{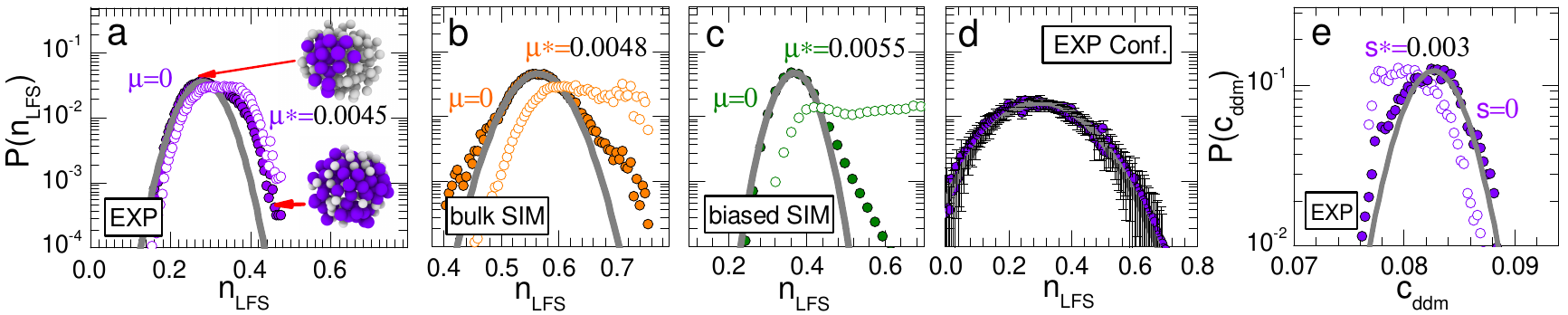}
\caption{Probability distributions of populations (filled symbols) of defective icosahedra in trajectories for the three systems we consider. Also shown is the post-processed, re-weighted data (open symbols) demonstrating coexistence between normal liquid and LFS-rich phases in each case. 
(a)~Experiment: sub-sampled, volume fraction $\phi=0.58$, trajectory length $t_\mathrm{obs}=0.97\tau_\alpha$. Post-processed data with $\mu^*=0.0045$.
(b)~Bulk simulation data for $N_\text{tot}=10960$ particles: sub-sampled, volume fraction $\phi=0.575$, $t_\mathrm{obs}=5\tau_\alpha$. Post-processed data with $\mu^*=0.0048$.
(c)~Biased simulation data for $N=125$ particles: full system with periodic boundaries, $\phi=0.56$, $t_\mathrm{obs}=10 \tau_\alpha$. Post-processed data with $\mu^*=0.0055$.
(d)~Confirming the transition is dynamical. Experimental probability distribution of defective icosahedra obtained from \emph{configurations} (rather than trajectories). 
(e)~Experiment: DDM data to show dynamical transition sub-sampled, $\phi=0.58$, $t_\mathrm{obs}=1\tau_\alpha$. In all panels, grey lines indicate Gaussian distributions and thus reveal large deviations [respectively their absence in (d)]. 
\paddyspeaks{Except where indicated, error bars are smaller than the symbols.}
} 
\label{figTrueMu}
\end{figure*}

\textit{Simulation and Analysis. --- }
We employ the DynamO event driven molecular dynamics package ~\cite{bannerman2011}. We consider a hard sphere system of five equimolar species of identical mass and different diameters :  $\{0.888, 0.9573, 1.0, 1.043, 1.112\}$. This system also has a polydispersity of 8\%. We fix the system size at $N=10976$. Timescales are scaled to the Brownian time of the experimental system. Further details can be found in the literature~\cite{royall2014,royall2014arxiv}. For the biased simulations, we follow the methods used previously~\cite{hedges2009,speck2012jcp,speck2012} with $N=125$ at $\phi = 0.56$. The trajectory length $t_\mathrm{obs}$ is chosen to be significantly greater than the relaxation time $t_\mathrm{obs} = 200 \simeq 10 \tau_\alpha$. Further details are discussed below and in the SM~\cite{SM}.

To analyze the local structure, we identify the bond network using the Voronoi construction with a maximum bond length of $1.4\sigma$.  \paddyspeaks{We then use the topological cluster classification (see SM) ~\cite{malins2013tcc}} to identify the locally favored structure for the hard spheres, the 10-membered defective icosahedron (an icosahedron missing three particles) with $C_{3v}$ symmetry depicted in  Fig. ~\ref{figAngell}(b)~\cite{royall2015}.

To determine the structural relaxation time $\tau_\alpha$ we calculate the intermediate scattering function (ISF) reading $F(t)=1/N\langle  \exp \left( {i\mathbf k\cdot[\mathbf r(t+t') - \mathbf r(t')]} \right)  \rangle$, where $|\mathbf{k}|=2 \pi$ is a wave-vector taken close to the peak of the static structure factor, $\mathbf r$ is the coordinate and the angle brackets indicate averaging over all particles. We do not discriminate between particles of different size here. The structural relaxation time is then obtained by fitting a stretched exponential $F(t)=c \exp \left[-(t/\tau_{\alpha})^{b} \right]$ as shown for experimental data in the SM~\cite{SM}. We compared experimental results with simulation through the Angell plot (Fig. \ref{figAngell}(a)), to obtain the effective volume fraction.

\textit{Overall system behaviour. --- } 
In Fig. \ref{figAngell}(a), we show the dynamical behaviour of the system where we plot the structural relaxation time against effective volume fraction for both experiments and simulations. Intermediate scattering functions are given in the SM \cite{SM}. We see that both experiments and simulations are well described by a Vogel-Fulcher-Tamman (VFT) fit $\tau_\alpha \propto \exp[A/(\phi_0-\phi)]$ in which $\phi_0=0.606t$ and $A=0.24$ parameterizes the fragility as shown in Fig. \ref{figAngell}(a), in line with previous work \cite{brambilla2009,royall2014}. In Fig. \ref{figAngell}(b), we see that upon increasing $\phi$, the population of locally favoured structures \cite{royall2014} increases both in experiment and simulation. Once the errors in coordinate tracking in the experiments are accounted for, we find quantitative agreement with simulation.

\textit{Evidence for a structural-dynamical phase transition. --- }
So far we have shown that the experimental hard sphere system undergoes structural change approaching dynamical arrest similar to the simulations \cite{royall2014}. Our strategy to provide evidence for a dynamical phase transition is as follows. First, we show that the hard sphere system undergoes the structural-dynamical phase transition previously identified in the 
\paddyspeaks{binary Lennard-Jones system} \cite{speck2012} in a small system of $N=125$ particles. We then proceed to show that the same behaviour, in the sense of a non-Gaussian probability distribution of the time-integrated fraction of particles in LFS, $n_\mathrm{LFS}$ is found in trajectories of $N=100$ particles which have been \emph{subsampled} from a bulk simulated system of $N_\text{tot}=10976$. This sets us up to perform a similar analysis on the experimental data. The larger than expected number of trajectories with a high population of LFS is then evidence for a dynamical phase transition in the experimental system. We then apply a bias through the dynamical chemical potential $\mu$ by post-processing unbiased simulated and experimental data, to reveal coexisting populations of normal liquid and LFS-rich phases.

\textit{Biased simulations. --- }
We compute the probability distribution for the population of LFS along trajectories, which is shown by the filled symbols in Fig. \ref{figTrueMu}(c). Here $\phi=0.56$.
We observe a peak at the equilibrium value $n_\mathrm{LFS} \simeq 0.37$, and a broad tail for high populations of LFS that differs significantly from the Gaussian distribution expected for normal liquids which are not supercooled/supersaturated. To bias the system towards phase coexistence between normal liquid and LFS rich phase, we promote those high population trajectories by reweighting the  $\mu = 0$ histogram.
\begin{equation}
\label{eq:pmu}
P_{\mu}(n_\mathrm{LFS}) \propto P(n_\mathrm{LFS})\,\exp[\mu n_\mathrm{LFS} N (K+1)]
\end{equation}
where $K+1$ is the number of frames in the trajectories. From the double-peaked distribution in Fig. \ref{figTrueMu}(c) we see that applying the $\mu$ field and increasing it above $\mu^* = 0.0055$ causes the system to undergo a transition from a low population of LFS ($n_\mathrm{LFS} \simeq 0.37$) to a high population $n_\mathrm{LFS} > 0.7$. By reweighting with $\mu^* = 0.0055$ we see that the tail rises to the same height of the first peak, indicating that at this value of the $\mu$ we have coexistence of the two phases in trajectory space. In other words, we have shown that hard spheres also exhibit the dynamical phase transition previously found
\cite{speck2012}.

\begin{figure}
\centering
\includegraphics[width=80mm]{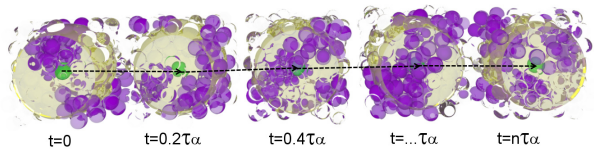} 
\caption{Illustration of the subsampling of trajectories. 
Purple particles are in defective icosahedra LFS, non-LFS are rendered transparent. In the experiments, we define the trajectories by considering the fraction of particles in LFS in a sphere which contains $\approx100$ particles (yellow tinted sphere). Here $n \tau_\alpha$ is the length of the trajectory.}
\label{figPurpleBalls}
\end{figure}

\textit{Bulk simulations. --- }
Having shown that the hard spheres undergo a structural-dynamical phase transition, we consider bulk simulations. Subsampled data for trajectories of $N=100$ particles and length $5\tau_\alpha$ are shown in Fig. \ref{figTrueMu}(b) for $\phi=0.575$. We subsample trajectories as shown schematically in Fig. \ref{figPurpleBalls}. In simulation the closest $N-1$ particles to a given particle define the trajectory. We 
harvest trajectories of length ${t_\mathrm{obs}=K{\bigtriangleup}t}$ , where ${K+1}$ is the total configuration by using 
$n_\mathrm{LFS}=\mathcal{N}/[N(K+1)]$ 
and ${\mathcal{N}}$ is the number of particles in LFS 
and ${\mathcal{N}=\displaystyle\sum_{k=0}^N\displaystyle\sum_{i=0}^K h^\mathrm{(LFS)}_k(t_i)}$.
Here $h^\mathrm{(LFS)}_k(t_i)=1$ if the particle is a member of an LFS and 0 otherwise.
Further details are shown in the SM \cite{SM}. We see that the trajectory distribution is again non-Gaussian and find a shoulder corresponding to LFS-rich trajectories, like the unbiased data in Fig. \ref{figTrueMu}(c) and 
that shown in ~\cite{speck2012}. 

\textit{Analysing unbiased trajectory data. --- }
The non-Gaussian behavior in Figs. \ref{figTrueMu}(b) and (c) with its characteristic ``fat tail'' demonstrates the dynamical phase transition. Here we go further to reveal phase coexistence by reweighting the trajectory distributions. To do so we apply the dynamical chemical potential $\mu$ via Eq.~(\ref{eq:pmu}). We see in Fig.~\ref{figTrueMu}(b) that applying the $\mu$ field leads to a distribution indicating the same two coexisting phases, identified under the biased simulations in Fig.~\ref{figTrueMu}(b), one LFS rich and one LFS poor (the normal liquid). Crucially, because we have subsampled from a large, \emph{unbiased} system we demonstrate that it is possible to identify the non-equlibrium phase transition in experimental data, which is itself of course unbiased.

\textit{Non-equilibrium phase transition in experiment. --- } We now proceed to demonstrate the non-equilibrium transition in experiment. Our strategy follows that applied to the large unbiased simulations above. In particular we subsampled the tracked coordinates from the experiment for trajectories of length $0.97\tau_\alpha$. For the experiments, trajectories are defined by the evolution of the $N-1$ closest particles assigned \emph{at the start} of the trajectory, see Fig.~\ref{figPurpleBalls} and the SM \cite{SM}. In our case  ${R\approx2.8\sigma}$, which corresponds to $\approx 100$ particles around a randomly chosen centre particle.

In Fig.~\ref{figTrueMu}(a) we plot the LFS trajectory distributions. As before, we see the characteristic non-Gaussian distribution of trajectories, indicating a non-equilibrium phase transition. We see similar behaviour to that of the simulations, in that there is a ``fat tail'' of LFS-rich trajectories, revealing the inactive phase. \paddyspeaks{Due to the particle tracking errors, the distribution has a lower mean in Fig.~\ref{figTrueMu}(a), however its relative width is comparable to that in Figs.~\ref{figTrueMu}(b) and (c).}

\begin{figure}[b]
\centering
\includegraphics[width=85mm]{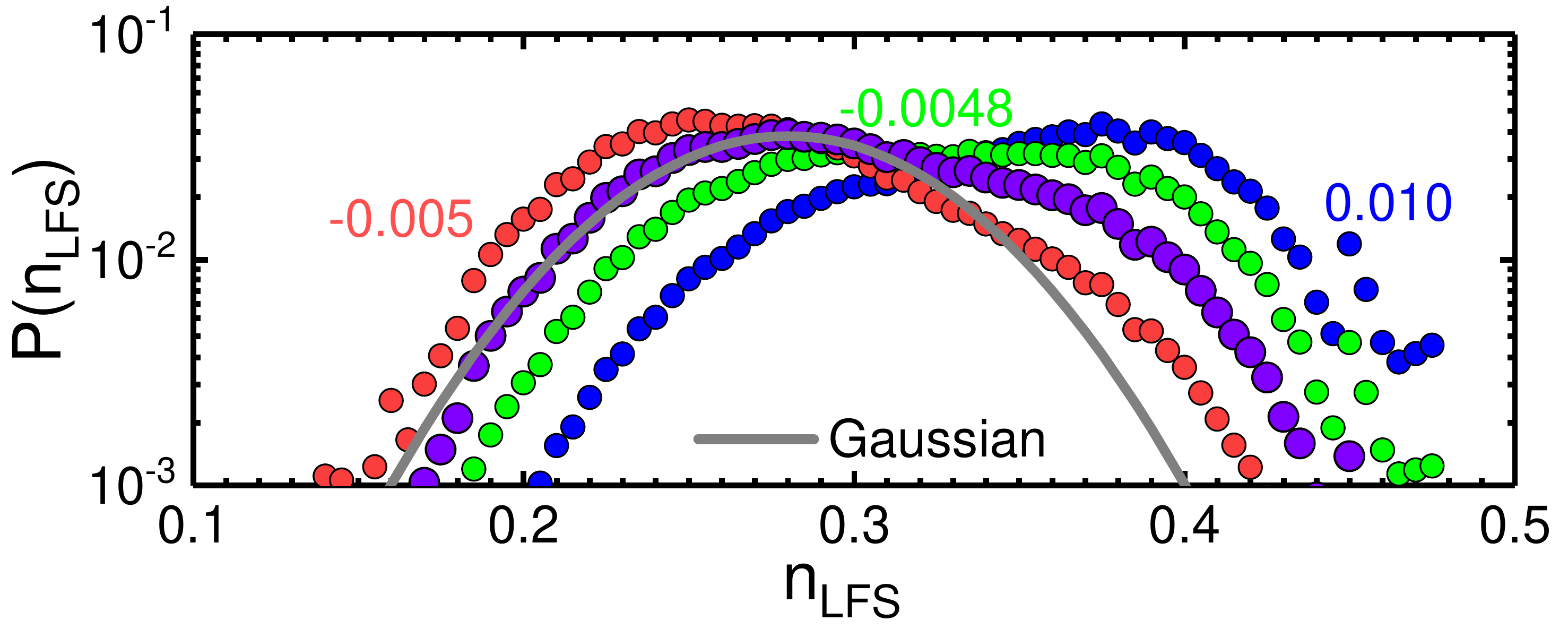}
\caption{Experimental probability distributions of \paddyspeaks{LFS} populations 
for several biases ${\mu}$ at volume fractions ~$\phi=0.58$.}
\label{figExpReweight}
\end{figure}

Significantly, we expect (as shown previously in simulation \cite{speck2012}), that simply sampling \emph{configurations} rather than \emph{trajectories}, then there should be a Gaussian distribution. That is to say, the phase transition has a dynamical character (rather than a conventional thermodynamic phase transition which would be revealed by coordinate data only). This we find, as shown in Fig. \ref{figTrueMu}(d). Thus we provide evidence that the transition is trajectory based, \emph{i.e.} that the dynamics are intrinsic to the transition, and thus it has a non-equilibrium nature. Another important check we need to make is that the transition is related to the particular LFS. In the SM \cite{SM} we show that trajectory sampling with a structure distinct from the LFS does not lead to a dynamical transition. Furthermore, we show that by controlling the dynamical chemical potential $\mu$, we can select either phase from the experimental data in Fig.~\ref{figExpReweight}. In this way it is possible, in experiment, to identify configurations of the inactive phase.

In Fig. \ref{figAngell}(b) (unfilled symbols), we estimate the volume fraction that the LFS-rich phase corresponds to as $0.59$. To do so, we determine the LFS population as a function of volume fraction $n_\mathrm{LFS}(\phi)$ (see SM \cite{SM}) as indicated by the grey lines in Fig. \ref{figAngell}(b). Under the VFT fit in Fig, \ref{figAngell}(a), this corresponds to a structural relaxation time 300 times that of the system from which the trajectories are sampled, $\phi=0.58$ for the experiments and some $1.8\times10^4$ in the case of the biased simulations, which are sampled at $\phi=0.56$.
In the future, with real-time data processing and using optical tweezers~\cite{williams2016} it may even be possible to ``freeze'' such an inactive configuration and further probe its behaviour, for example by determining its rheological properties.

Finally we consider the purely dynamical transition to a state of trajectories with very slow dynamics. This is shown in Fig. \ref{figTrueMu}(e). \paddyspeaks{Now the measurements of the displacements necessary are rather hampered by the particle tracking errors. We therefore }
determine mobility with confocal 
differential dynamic microscopy (ConDDM) \cite{lu2012, cerbino2008} as described in the SM \cite{SM}. We see that there is a ``fat tail'' for low mobility indicating a dynamical transition. This is also found in simulation, for which details are presented in the SM \cite{SM}.

\textit{Conclusions. --- }
We have demonstrated the existence of a dynamical phase transition in trajectory space \emph{in experiment} between a normal liquid and an LFS-rich phase. This opens a perspective as to the range of dynamical phase transitions that might be identified by this kind of analysis. Here we have focused mainly on structure (which is easier to identify in \paddyspeaks{our} experiments
) but have also demonstrated the purely dynamical phase transition. We have previously shown that there appears to be some overlap between the configuration space these transitions sample \cite{speck2012}. We see no reason to suppose that the current hard spheres should be significantly different. While some work has suggested that the hard sphere LFS might have a hexgaonal symmetry \cite{tanaka2010}, no evidence of such order has been seen in a number of other studies, including this \cite{charbonneau2012,royall2015,royall2014arxiv}. \paddyspeaks{Finally, we find that trajectory biasing based on LFS can produce configurations of exceptionally low configurational entropy, suggesting a link between LFS and configurational entropy \cite{turci2016}.}

\begin{acknowledgments}
The authors are grateful to M. Leocmach for his generous help with data analysis and to Peter Crowther for assistance with the DDM method. CPR gratefully acknowledges the Royal Society and CPR, JH and FT European Research Council (ERC Consolidator Grant NANOPRS, project number 617266) for financial support. RP thanks Development and Promotion of Science and Technology Talented Project (DPST) for financial support. MC is funded by the DFG through the Graduate School ``Materials Science in Mainz'' (GSC 266). Some of this work was carried out using the facilities of the Advanced Computing Research Centre, University of Bristol. CPR acknowledges the University of Kyoto SPIRITS fund.
\end{acknowledgments}

\nocite{leocmach2013sm}
\nocite{royall2016}
\nocite{poon2012}
\nocite{berthier2007}
\nocite{thorneywork2015}
\nocite{bolhuis2002}
\nocite{minh2009}

\bibliographystyle{apsrev}
\bibliography{trueMu}

\end{document}